\documentclass{svjour3a}                     
\smartqed  
\usepackage{graphicx}
\usepackage{mathptmx}      
\usepackage{enumerate}
\usepackage{dsfont}

\usepackage[english]{babel}
\usepackage[latin1]{inputenc}
\usepackage{times}
\usepackage{fixmath}
\usepackage{amssymb,amsfonts,amsmath}

\DeclareSymbolFont{extra}{OML}{cmm}{m}{n}
\DeclareMathSymbol{\varrho}{\mathord}{extra}{'045} 
\DeclareMathSymbol{\nu}{\mathord}{extra}{'027}
\DeclareMathSymbol{\zeta}{\mathord}{extra}{'020}
\DeclareMathSymbol{\kappa}{\mathord}{extra}{'024}
\DeclareMathSymbol{\omega}{\mathord}{extra}{'041}
\DeclareMathSymbol{\Ph}{\mathord}{extra}{'10}
\DeclareMathSymbol{\Omega}{\mathord}{extra}{'12}
\DeclareMathSymbol{\alpha}{\mathord}{extra}{'13}
\DeclareMathSymbol{\beta}{\mathord}{extra}{'14}
\DeclareMathSymbol{\gamma}{\mathord}{extra}{'15}
\DeclareMathSymbol{\delta}{\mathord}{extra}{'16}
\DeclareMathSymbol{\eta}{\mathord}{extra}{'21}
\DeclareMathSymbol{\xi}{\mathord}{extra}{'30}
\DeclareMathSymbol{\varepsilon}{\mathord}{extra}{'42}
\DeclareMathSymbol{\varphi}{\mathord}{extra}{'47}
\DeclareMathSymbol{\pi}{\mathord}{extra}{'31}
\DeclareMathSymbol{\phi}{\mathord}{extra}{'36}
\DeclareMathSymbol{\theta}{\mathord}{extra}{'22}

\newcommand{\dd}{\mathrm{d}}
\newcommand{\ee}{\mathrm{e}}
\newcommand{\ii}{\mathrm{i}}
\newcommand{\R}{\mathds R}
\newcommand{\C}{\mathds C}
\newcommand{\eps}{\varepsilon}

\newcommand{\rW}{\varrho}
\newcommand{\zW}{\zeta}

\newcommand{\Ap}{\mathcal A^+}
\newcommand{\An}{\mathcal A^0}
\newcommand{\Am}{\mathcal A^-}
\newcommand{\Ha}{\mathcal{H}_1}
\newcommand{\Hb}{\mathcal{H}_2}
\newcommand{\bPhi}{{\mathbold\Phi}}

 \journalname{General Relativity and Gravitation}
\begin{document}

\title{Non-existence of stationary two-black-hole configurations: The
degenerate case}
\author{J\"org Hennig \and Gernot Neugebauer}


\institute{J. Hennig \at
           Department of Mathematics and Statistics,
           University of Otago,
           P.O. Box 56, Dunedin 9054, New Zealand
           \email{jhennig@maths.otago.ac.nz}
           \and
           G. Neugebauer\at
          Theoretisch-Physikalisches Institut,
          Friedrich-Schiller-Universit\"at,
          Max-Wien-Platz 1,
          07743 Jena, Germany\\
          \email{G.Neugebauer@tpi.uni-jena.de}       
}

\date{{\bf Manuscript date: \today}}

\maketitle
\begin{abstract}

In a preceding paper we examined the question whether the spin-spin repulsion and the gravitational attraction of two aligned sub-extremal black holes can balance each other. Based on the solution of a boundary value problem for two separate (Killing-) horizons and a novel black hole criterion we were able to prove the non-existence of the equilibrium configuration in question. In this paper we extend the non-existence proof to extremal black holes.

\end{abstract}
\section{Introduction}

Our intuition tells us that \emph{static} equilibrium configurations consisting of gravitationally interacting bodies at rest cannot exist. In fact advancement has recently been made on the way to a general non-existence proof \cite{Beig1}. For a corresponding configuration of \emph{rotating} bodies the problem is less transparent. It is imaginable that the interaction of the angular momenta (``spin-spin interaction'') could generate repulsive effects compensating the omnipresent mass attraction. 

In a preceding paper \cite{Neugebauer2009} (henceforth denoted as paper I) we discussed, as a characteristic example, this question for two aligned rotating black holes and arrived at a negative conclusion: For two sub-extremal black holes the anticipated equilibrium configuration does not exist. For special symmetric configurations consisting of rotating bodies a similar result was obtained in \cite{Beig2}.

This paper is meant to extend the discussion to extremal (degenerate) black holes. Besides the mathematical completeness, it is physically interesting to study objects with high angular momenta.

\begin{figure}\centering
 \includegraphics[width=0.999\textwidth]{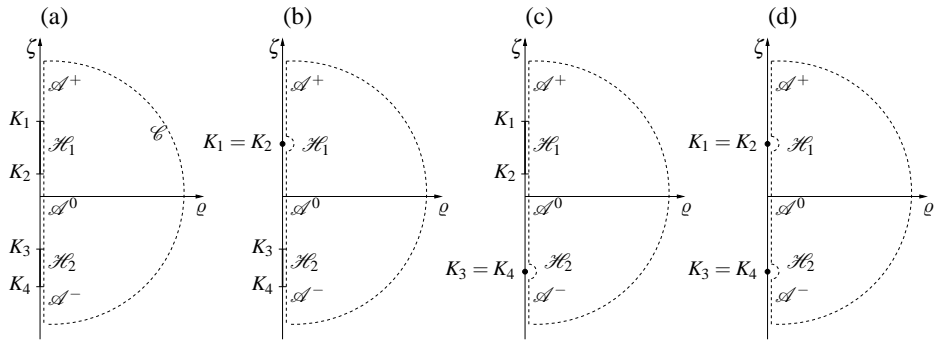}
 \caption{Illustration of the two-black-hole equilibrium
 configurations in Weyl coordinates. The event
 horizons $\mathcal H_1$ and $\mathcal H_2$ of the two black holes
 are located in the
 intervals $[K_2,K_1]$ and $[K_4,K_3]$ on the $\zW$-axis, respectively.
 The remaining parts $\mathcal A^\pm$, $\mathcal A^0$
 of the $\zW$-axis correspond to the rotation axis. Configurations with
 two sub-extremal black holes (a) have been investigated in
 \cite{Neugebauer2009}. Here we focus on configurations containing
 degenerate black holes. The three different types of configurations with degenerate (``point-like'')  horizons are sketched in (b), (c), (d).}
 \label{fig:1}
\end{figure}

Again (see paper I), we want to emphasize that a discussion like this cannot be based on an arbitrarily chosen solution of Einstein's field equations (``double-Kerr-NUT'' etc.). Instead we follow the idea of paper I and formulate a boundary value problem for two separate (Killing) horizons (see Fig.~\ref{fig:1}). 
Since the problem is stationary and axisymmetric we can apply the mathematical tools developed in the context of the Ernst equation. In particular, we use Weyl-Lewis-Papapetrou coordinates\footnote{As was shown by Chru\'sciel and Costa \cite{Chrusciel1},
Weyl-Lewis-Papapetrou coordinates can indeed be
introduced as \emph{global} coordinates in the axisymmetric, stationary and asymptotically flat vacuum region outside of rotating black holes. This is true even in the degenerate case
\cite{Chrusciel2}.} and calculate the Ernst potential on the axis of symmetry by applying the inverse scattering technique. As a peculiarity of the Weyl-Lewis-Papapetrou coordinates, the horizons of black holes are located on the $\zW$-axis. While sub-extremal horizons cover
intervals $K_1\ge\zW\ge K_2$ or/and $K_3\ge\zW\ge K_4$, see Figs.~\ref{fig:1}a-c, \emph{extreme} horizons are represented by points $\zW=K_1=K_2$ or/and $\zW=K_3=K_4$, see Figs.~\ref{fig:1}b-d. In all cases, the Ernst potential on the regular parts of the $\zW$-axis $\Ap$, $\Am$, $\An$ turns out to be the quotient of two normalized polynomials of the second degree in $\zW$, whose coefficients have to satisfy a set of algebraic conditions expressing the regularity of the metric on $\Ap$, $\Am$, $\An$ . Finally, the analysis of these conditions together with other physical restrictions, such as the positiveness of the mass of the system, a specific black hole inequality connecting angular momentum and horizon area or the exclusion of singular rings, leads to a non-existence theorem for two-black-hole configurations.
Since the sub-extremal case (Fig.~\ref{fig:1}a) was already discussed in paper I, we will here focus our attention on configurations with ``point-like'' horizons (Figs.~\ref{fig:1}b-d).

\section{The boundary value problem}

Following paper I we describe the gravitational vacuum fields outside the horizons in cylindrical Weyl-Lewis-Papapetrou coordinates $(\rW,\zW,\varphi,t)$ by the line element

\begin{equation}\label{LE}
 \dd s^2 = \ee^{-2U}\big[\ee^{2k}(\dd\rW^2+\dd\zW^2)
           +\rW^2\dd\varphi^2\big]
           -\ee^{2U}(\dd t+a\,\dd\varphi)^2,
\end{equation}
where the Newtonian gravitational potential $U$, the gravitomagnetic potential $a$ and the superpotential $k$ are functions of $\rW$ and $\zW$ alone. Hence the metric \eqref{LE} admits an Abelian group of motions $G_2$ with the generators (Killing vectors)
\begin{equation}
\begin{aligned}
 &\xi^i=\delta^i_t,       &\xi^i \xi_i<0 & \quad\textrm{(stationarity)}\\
 &\eta^i=\delta^i_\varphi,&\eta^i\eta_i>0 & \quad\textrm{(axisymmetry)},
\end{aligned}
\end{equation}
where the Kronecker symbols $\delta^i_t$, $\delta^i_\varphi$ indicate that $\xi^i$ has only a $t$-component whereas $\eta^i$ points in the azimuthal ($\varphi$-) direction along closed circles. (Note that $\eta^i\eta_i>0$ must hold everywhere off the symmetry axis, whereas $\xi^i\xi_i<0$ is assumed only sufficiently far away from the two black holes. Indeed, in the interior of possible ergoregions around the black holes, $\xi^i$ would become spacelike.) Obviously,
\begin{equation}
 \ee^{2U}=-\xi^i\xi_i,\quad a=-\ee^{-2U}\eta^i\xi_i
\end{equation}
is a coordinate-free representation of the gravitational potentials $U$ and $a$.

In Weyl-Lewis-Papapetrou coordinates, event horizons degenerate to one-dimensional pieces of the $\zW$-axis, see, e.g. Fig.~\ref{fig:1}a with the horizons $\Ha:\,\rW=0,\, K_1\ge\zW\ge K_2$; $\Hb:\,\rW=0,\, K_3\ge\zW\ge K_4$ or isolated points, see Figs.~\ref{fig:1}b-d. While we discussed the configurations \ref{fig:1}a with two extended (``sub-extremal'') horizons $\Ha$, $\Hb$ in paper I, this paper deals with the degenerate cases \ref{fig:1}b-d. Again, we can make use of the fact that event horizons in stationary and axisymmetric spacetimes are \emph{Killing horizons}, see paper I. 

A Killing horizon can be defined by a linear combination $L$ of the Killing vectors $\xi$ and $\eta$,
\begin{equation}
 L=\xi+\Omega\eta
\end{equation}
with the norm
\begin{equation}\label{Vdef}
 \ee^{2V}:=-(L,L)=\ee^{2U}\left[(1+\Omega a)^2-\rW^2\Omega^2\ee^{-4U}\right]
\end{equation}
where $\Omega$ is the constant angular velocity of the horizon. A connected component of the set of points with $\ee^{2V}=-(L,L)=0$, which is a null hypersurface, $(\dd\ee^{2V},\dd\ee^{2V})=0$, is called a Killing horizon $\mathcal H(L)$,
\begin{equation}\label{KH}
  \mathcal H(L):\quad \ee^{2V}=-(L,L)=0,\quad (\dd\ee^{2V},\dd\ee^{2V})=0.
 \end{equation}
 
 Since the Lie derivative $\mathcal L_L$ of $\ee^{2V}$ vanishes, we have $(L,\dd\ee^{2V})=0$. Being null vectors on $\mathcal H(L)$, $L$ and $\dd\ee^{2V}$ are proportional to each other,
\begin{equation}\label{kappa1}
 \mathcal H(L):\quad \dd\ee^{2V}=-2\kappa L.
\end{equation}
Using the field equations one can show that the \emph{surface gravity} $\kappa$ is a constant on $\mathcal H(L)$. In Weyl-Lewis-Papapetrou coordinates, the event horizon degenerates to a ``straight line'' and covers an interval $(\zW_1,\zW_2)$ of the $\zW$-axis, $\rW=0$ (``extended case'', see paper I) or shrinks to a single point $\zW=\zW_0$, $\rW=0$ (``extreme case'', subject of this paper). Note that a Killing horizon is nonetheless a two-surface: The degeneracy to a line or a point is a peculiarity of the Weyl-Lewis-Papapetrou coordinate system.

In the subsequent discussion we will essentially use the Ernst formulation of the field equations. For that purpose, we introduce the complex \emph{Ernst potential}
\begin{equation}
 f=\ee^{2U}+\ii b,
\end{equation}
where the \emph{twist potential} $b$ is defined in terms of the metric
potential $a$ via
\begin{equation}\label{a}
 a_{,\rW} = \rW\,\ee^{-4U} b_{,\zW},\qquad
 a_{,\zW} = -\rW\,\ee^{-4U} b_{,\rW}.
\end{equation}
In this formulation, the Einstein vacuum equations are equivalent to the
complex \emph{Ernst equation}
\begin{equation}\label{Ernst}
 (\Re f)\Big(f_{,\rW\rW}+f_{,\zW\zW} +\frac{1}{\rW}f_{,\rW}\Big)
 = f_{,\rW}^2 + f_{,\zW}^2.
\end{equation}
The metric potential $k$ can be calculated from $f$ via a line integral,
\begin{equation}
 \label{k1}
 k_{,\rW} = \rW\Big[U_{,\rW}^2-U_{,\zW}^2+\frac{1}{4}\ee^{-4U}
         (b_{,\rW}^2-b_{,\zW}^2)\Big],\quad
 k_{,\zW} = 2\rW\Big[U_{,\rW}U_{,\zW}+\frac{1}{4}\ee^{-4U}
         b_{,\rW}b_{,\zW}\Big].
\end{equation}

Fig.~\ref{fig:1} sketches the boundaries of the vacuum region: $\Ap$, $\An$, $\Am$ are the regular parts of the $\zW$-axis, $\Ha$ and $\Hb$ denote the Killing horizons of the two black holes and $\mathcal C$ stands for spatial infinity. 
In a first step of the non-existence proof we will solve the Ernst equation under the boundary conditions 
\begin{eqnarray}
 \label{B1}
 \mathcal A^\pm,\An: && \quad a=0,\quad k=0,\\
 \label{B2}
 \mathcal H_i:       && \quad 1+\Omega_i a=0,\quad i=1,2,\\
 \label{B3}
 \mathcal C:         && \quad U\to 0,\quad a\to 0,\quad k\to 0.          
\end{eqnarray}
where $\Omega_1$ and $\Omega_2$ are the angular velocities of the two
horizons. The first equation expresses characteristics of the axis of symmetry. The second equation reflects the attribute $\ee^{2V}=0$, $\rW=0$ of Killing horizons, see
\eqref{Vdef} and the third equation ensures asymptotic flatness of the four-metric \eqref{LE}. Equations \eqref{B2} and \eqref{kappa1} tell us that the anticipated two-black-hole solution will depend on the four characteristic ``horizon constants'' $\kappa_1$, $\Omega_1$ and $\kappa_2$, $\Omega_2$. One can show that 
\begin{equation}\label{kapom}
 \kappa_1+\ii\Omega_1=\frac{1}{2} f^+_{,\zW}\big|_{\zW=K_1}, \quad
 \kappa_2+\ii\Omega_2=\frac{1}{2} f^0_{,\zW}\big|_{\zW=K_3},
\end{equation}
where $f^+$ and $f^0$ are the axis values of the Ernst potential on $\Ap$ and $\An$, respectively. The proof of these relations makes use of the fact that the vectors $L_{(i)}=\xi+\Omega_i\eta$ $(i=1,2)$ are defined everywhere on and inside the boundaries. The constancy of $\kappa_i$, $\Omega_i$ on the respective horizon implies
\begin{equation}\label{B4}
  f^+_{,\zW}\big|_{\zW=K_1}= \bar f^0_{,\zW}\big|_{\zW=K_2},\quad
  f^0_{,\zW}\big|_{\zW=K_3}= \bar f^-_{,\zW}\big|_{\zW=K_4},
\end{equation}
where the bar denotes complex conjugation. The conditions \eqref{B1}-\eqref{B3} and 
\eqref{B4} are not independent of each other. However, we need not discuss this interrelationship since the Eqs. \eqref{B1}-\eqref{B3} alone are necessary conditions and will lead to a solution of the Ernst equation with inevitable defects. 

\section{Solution with the inverse scattering method}
\subsection{The linear problem}

As was shown in \cite{Neugebauer2000,Neugebauer2003}, the boundary value
problem for two axisymmetric and stationary black holes can be solved
with the \emph{inverse scattering method} --- a powerful technique coming
from soliton theory. Hereby, an associated linear problem (LP) is analyzed,
whose integrability condition is equivalent to the nonlinear Ernst
equation.

We use the LP
\cite{Neugebauer1979,Neugebauer1980b}
\begin{equation}\label{LP}
\begin{aligned}
 \bPhi_{,z} & = \left[\left(\begin{array}{cc}
                  B & 0\\
                  0 & A\end{array}\right)
                  +\lambda\left(\begin{array}{cc}
                  0 & B\\
                  A & 0\end{array}\right)\right]\bPhi,\\
 \bPhi_{,\bar z} & = \left[\left(\begin{array}{cc}
                  \bar A & 0\\
                  0 & \bar B\end{array}\right)
                  +\frac{1}{\lambda}\left(\begin{array}{cc}
                  0 & \bar A\\
                  \bar B & 0\end{array}\right)\right]\bPhi,
\end{aligned}                 
\end{equation}
where the \emph{pseudopotential}
$\bPhi(z,\bar z,\lambda)$ is a $2\times2$ matrix depending on the
spectral parameter
\begin{equation}\label{lambda}
 \lambda=\sqrt{\frac{K-\ii\bar z}{K+\ii z}},\quad K\in\C,
\end{equation}
as well as on the complex coordinates
\begin{equation}
 z=\rW+\ii\zW,\quad \bar z=\rW-\ii\zW,
\end{equation}
whereas
\begin{equation}
 A=\frac{f_{,z}}{f+\bar f},\qquad
 B=\frac{\bar f_{,z}}{f+\bar f}
\end{equation}
and the complex conjugate quantities $\bar A$, $\bar B$ are functions of
$z$, $\bar z$ (or $\rW$, $\zW$) alone and do not depend on the constant
parameter $K$.

The idea of the inverse scattering method is to construct $\bPhi$, for
fixed but arbitrary values of $z$, $\bar z$, as a holomorphic function
of $\lambda$ and to calculate the Ernst potential $f(\rW,\zW)$ via
\begin{equation}
 f(\rW,\zW)=\bPhi_{21}(z,\bar z,1)
\end{equation}
from $\bPhi$. To do so, we have to distinguish between situations with
``extended horizons'' (the horizons are intervals with positive lengths,
i.e. $K_1\neq K_2$ and $K_3\neq K_4$) and situations with ``point-like
horizons'' ($K_1=K_2$ and/or $K_3=K_4$).

The construction of $\bPhi$ starts with an integration of the LP along the closed dashed lines as sketched in Figs.~\ref{fig:1}a-d which embrace the domain outside the horizons. The procedure makes explicit use of the boundary conditions \eqref{B1}-\eqref{B3} and simplifications of the differential equations \eqref{LP}  due to $\lambda=\pm 1$ for $\rW=0$ and $A=0=B$ along $\mathcal C$. For a detailed description of the single steps in the case of two \emph{extended} horizons (Fig.~\ref{fig:1}a, $K_1>K_2>K_3>K_4$) see \cite{Neugebauer2000,Neugebauer2003} and paper I. The result is a matrix representation $\mathcal N$ of the axis values of the Ernst potential on $\Ap$, $f^+(\zW)\equiv f^+(\rW=0,\zW)=\ee^{2U^+(\zW)}+\ii b^+(\zW)$,
\begin{equation}\label{Nmat1}
 \mathcal N := \ee^{-2U^+(\zW)}\left(\begin{array}{cc}
               1 & -\ii b^+(\zW)\\[1ex]
               \ii b^+(\zW) & \  f^+(\zW)\bar f^+(\zW)
               \end{array}\right)
\end{equation}
in terms of the parameters $K_n$, $f_n=f(\rW=0,\zW=K_n)$, $n=1,\dots,4$, and the angular velocities $\Omega_1=\Omega^{(1)}=\Omega^{(2)}$, $\Omega_2=\Omega^{(3)}=\Omega^{(4)}$,
\begin{equation}\label{Nmat}
 \mathcal N = \prod\limits_{n=1}^4\left({\bf 1}
               +\frac{{\bf F}_n}{2\ii\Omega^{(n)}(\zW-K_n)}\right),
\end{equation}
where
\begin{equation}\label{defF}
 {\bf F}_n :=(-1)^n\left(\begin{array}{cc}
             f_n & -1\\[1ex]
             f_n^2 & \  -f_n
             \end{array}\right).
\end{equation}
(Note that \eqref{Nmat} is only valid for $\Omega_1\neq0\neq\Omega_2$. However, the case of vanishing angular velocities can be treated similarly by starting the entire discussion in a rotating frame of reference. In this way, one obtains a formula similar to \eqref{Nmat} which also leads to the main result \eqref{fp} below.)

According to $\mathcal N$ in \eqref{Nmat1}, the sum of the off-diagonal elements has to vanish, $\mathcal N_{12}+\mathcal N_{21}=0$, whence
\begin{equation}\label{constraint}
 \mathrm{tr} \left[\left(\begin{array}{cc}
            0 & 1\\ 1 & 0\end{array}\right)
             \prod\limits_{n=1}^4\left(
             {\bf 1}+\frac{{\bf F}_n}{2\ii\Omega^{(n)}(\zW-K_n)}\right)
             \right] = 0.
\end{equation}
Since this equation holds identically in $\zW$, one obtains four
restrictions among $\Omega_1$, $\Omega_2$; $K_3-K_4$, $K_2-K_3$,
$K_1-K_2$; $f_1$,\dots $f_4$. The discussion of the restrictions for extended horizons was an important step towards the non-existence proof in paper I. To repeat the integration of the LP for point-like horizons, on has to switch over to appropriate coordinates, mapping the horizon on a one-dimensional domain. As was shown by Meinel, see \cite{Meinel2008}, for a single extreme black hole at $\rW=0$, $\zW=K_0$ a suitable transformation is
\begin{equation}
 \rW=R\sin\theta,\quad \zW=K_0+R\cos\theta.
\end{equation}
In the new coordinates $R$, $\theta$, the horizon is described by $R\to 0$, $\theta\in[0,\pi]$, i.e. by a line in an $R$-$\theta$-diagram. Applying this idea to the ``point-like'' configurations as sketched in Figs.~\ref{fig:1}b-d and performing the integration along the boundaries (dashed lines) one again arrives at \eqref{Nmat}-\eqref{constraint}, in which one simply has to put $K_1=K_2$ (Fig.~\ref{fig:1}b), $K_3=K_4$ (Fig.~\ref{fig:1}c) or $K_1=K_2$ and $K_3=K_4$ (Fig.~\ref{fig:1}d). 

In all cases (Figs.~\ref{fig:1}a-d), the restriction \eqref{constraint} tells us that the Ernst potential $f^+(\zW)$ is a quotient of two normalized polynomials of second degree,
\begin{equation}\label{fp}
 f^+(\zW)=\frac{n_2(\zW)}{d_2(\zW)}\equiv
 \frac{\zW^2+q\zW+r}{\zW^2+s\zW+t},\quad
 q,r,s,t\in\C.
\end{equation}
The proof of \eqref{fp} follows the argumentation in paper I which is still valid for point-like horizons. 

In order to prepare the continuation of $f^+(\zW)$ off the axis of symmetry into the $\rW$-$\zW$-plane, we replace the constants $q$, $r$, $s$, $t$ by appropriate parameters. For this reason we introduce the functions
\begin{equation}\label{albet}
 \alpha(\zW)=\frac{\bar d_2(\zW)}{d_2(\zW)},\quad \alpha\bar\alpha=1,\quad
 \beta(\zW)=\frac{\bar n_2(\zW)}{n_2(\zW)},\quad \beta\bar\beta=1
\end{equation}
and investigate their behaviour at the constant parameters $K_i$ ($i=1,\dots,4$), fixing the positions of the horizons, see Figs.~\ref{fig:1}a-d. Obviously, the treatment of configurations with point-like horizons (Figs.~\ref{fig:1}b-d) will differ from the treatment of the ``extended'' case (Fig.~\ref{fig:1}a).

\subsection{Extended horizons}

In this section we summarize the results of paper I.

Introducing the parameters
\begin{equation}\label{albet1}
\begin{aligned}
 &\alpha_i=\alpha(K_i),& \alpha_i\bar\alpha_i=1,\quad i=1,\dots,4,\\
 &\beta_i=\beta(K_i),& \beta_i\bar\beta_i=1,\quad i=1,\dots,4,
\end{aligned}
\end{equation}
in \eqref{albet}, we obtain the two linear algebraic systems of equations
\begin{equation}\label{lineq}
 \bar d_2(K_i)-\alpha_i d_2(K_i)=0,\quad
 \bar n_2(K_i)-\beta_i n_2(K_i)=0, \quad i=1,\dots,4
\end{equation}
for $s$, $t$ ($\bar s$, $\bar t$); $q$, $r$ ($\bar q$, $\bar r$) with $K_i$, $\alpha_i$, $\beta_i$ as coefficients. According to \eqref{Nmat1}-\eqref{defF} we have
\begin{equation}\label{Up}
 \ee^{2U^+}=\mathcal N^{-1}_{11}
 =\frac{(\zW-K_1)(\zW-K_2)(\zW-K_3)(\zW-K_4)}{p_4(\zW)}, 
\end{equation}
where $p_4(\zW)$ is a real normalized polynomial of the fourth degree in $\zW$. From $\ee^{2U^+(K_i)}=0$ ($p_4(K_i)\neq 0$, $i=1,\dots, 4$) we get
\begin{equation}
 \bar f^+(K_i)=-f^+(K_i),\quad i=1,\dots,4
\end{equation}
whence
\begin{equation}
 \beta_i=-\alpha_i.
\end{equation}

Hence, $f^+$ can be expressed in terms of $\alpha_i$ (and $K_i$) alone. Solving the linear equations \eqref{lineq} for $q$, $r$, $s$, $t$ and plugging the result into \eqref{fp} we arrive at a determinant representation of the axis potential $f^+$ on $\Ap$,
\begin{equation}\label{axisf}
 f^+(\zW) = \frac{\left|\begin{array}{ccccc}
   1 & K_1^2  & K_2^2  & K_3^2 & K_4^2\\[1ex]
  \ 1\ & \ \alpha_1K_1(\zW-K_1)\ & \ \alpha_2K_2(\zW-K_2)\ &
   \ \alpha_3K_3(\zW-K_3)\ & \ \alpha_4K_4(\zW-K_4)\ \\[1ex]
   0 & K_1    & K_2    & K_3   & K_4\\[1ex]
   0 & \alpha_1(\zW-K_1) & \alpha_2(\zW-K_2) &
   \alpha_3(\zW-K_3) & \alpha_4(\zW-K_4)\\[1ex]
   0 & 1      & 1      & 1     & 1             
   \end{array}\right|}
   {\left|\begin{array}{ccccc}
   1 & K_1^2  & K_2^2  & K_3^2 & K_4^2\\[1ex]
  -1 & \ \alpha_1K_1(\zW-K_1)\ & \ \alpha_2K_2(\zW-K_2)\ &
   \ \alpha_3K_3(\zW-K_3)\ & \ \alpha_4K_4(\zW-K_4)\ \\[1ex]
   0 & K_1    & K_2    & K_3   & K_4\\[1ex]
   0 & \alpha_1(\zW-K_1) & \alpha_2(\zW-K_2) &
   \alpha_3(\zW-K_3) & \alpha_4(\zW-K_4)\\[1ex]
   0 & 1      & 1      & 1     & 1             
   \end{array}\right|}.
\end{equation}
It can easily be seen that
\begin{equation}\label{genf}
 f(\rW,\zW) = \frac{\left|\begin{array}{ccccc}
   1 & K_1^2  & K_2^2  & K_3^2 & K_4^2\\[1ex]
  \ 1\ & \ \alpha_1K_1r_1\ & \ \alpha_2K_2r_2\ &
   \ \alpha_3K_3r_3\ & \ \alpha_4K_4r_4\ \\[1ex]
   0 & K_1    & K_2    & K_3   & K_4\\[1ex]
   0 & \alpha_1r_1 & \alpha_2r_2 &
   \alpha_3r_3 & \alpha_4r_4\\[1ex]
   0 & 1      & 1      & 1     & 1             
   \end{array}\right|}
   {\left|\begin{array}{ccccc}
   1 & K_1^2  & K_2^2  & K_3^2 & K_4^2\\[1ex]
  -1 & \ \alpha_1K_1r_1\ & \ \alpha_2K_2r_2\ &
   \ \alpha_3K_3r_3\ & \ \alpha_4K_4r_4\ \\[1ex]
   0 & K_1    & K_2    & K_3   & K_4\\[1ex]
   0 & \alpha_1r_1 & \alpha_2r_2 &
   \alpha_3r_3 & \alpha_4r_4\\[1ex]
   0 & 1      & 1      & 1     & 1             
   \end{array}\right|},
\end{equation}
where
\begin{equation}\label{ri}
 r_i := \sqrt{(\zW-K_i)^2+\rW^2}\ge0,\qquad i=1,\dots,4,
\end{equation}
is a continuation of $f^+(\zW)$ to all space. 
A concise reformulation of this expression originates from Yamazaki \cite{Yamazaki},
\begin{equation}\label{Yamazaki}
 f(\rW,\zW) = \frac{\left|\begin{array}{cc}
              R_{12}-1 & R_{14}-1\\
              R_{23}-1 & R_{34}-1\end{array}\right|}
              {\left|\begin{array}{cc}
              R_{12}+1 & R_{14}+1\\
              R_{23}+1 & R_{34}+1\end{array}\right|},\qquad
 R_{ij}:=\frac{\alpha_ir_i-\alpha_jr_j}{K_i-K_j}. 
\end{equation}
As was shown in \cite{Kramer1980,Neugebauer1980a}, $f(\rW,\zW)$ in \eqref{genf} is the Ernst potential of the double-Kerr-NUT solution, and, as a solution of the Ernst equation uniquely determined by its axis values $f^+(\zW)$. Thus, all solutions of the balance problem for two \emph{extended} horizons are necessarily contained in the double Kerr-NUT family of solutions of the Einstein vacuum equations, see \cite{Kramer1980}. Applying the boundary conditions \eqref{B1} and \eqref{B3} to this solution, Tomimatsu and Kihara \cite{Tomimatsu} derived a complete set of algebraic equilibrium conditions on the axis of symmetry between the parameters $\alpha_i$, $K_i$ ($i=1,\dots,4$) which were explicitly solved by Manko \emph{et al.} \cite{Manko2000}. It turns out that the only condition to satisfy $k=0$ on $\mathcal A^\pm$, $\An$ is
\begin{equation}\label{Bed1}
 \alpha_1\alpha_2+\alpha_3\alpha_4=0.
\end{equation}
Combining this result with $a^\pm=0$, $a^0=0$ ($a(\rW,\zW)$ again calculated from $f(\rW,\zW)$, see \eqref{a}), one obtains two further conditions,

\begin{equation}\label{Bed2}
\begin{aligned}
 \frac{(1-\alpha_4)^2}{\alpha_4}w^2&=\frac{(1-\alpha_3)^2}{\alpha_3},\quad
   &w^2&:=\frac{K_{14}K_{24}}{K_{13}K_{23}}\in[1,\infty),\\ 
 \frac{(1+\alpha_2)^2}{\alpha_2}w'^2&=\frac{(1+\alpha_1)^2}{\alpha_1},\quad
   &w'^2&:=\frac{K_{23}K_{24}}{K_{13}K_{14}}\in(0,1],
\end{aligned}
\end{equation}
where
\begin{equation}
 K_{ij}:=K_i-K_j.
\end{equation}

One can show that the restrictions \eqref{constraint} are identically satisfied if the conditions \eqref{Bed1} and \eqref{Bed2} hold. This may be proved using the explicit solution of \eqref{Bed1}, \eqref{Bed2},
\begin{equation}\label{alpha1}
 \begin{aligned}
  &\alpha_1 = \frac{w'\alpha^2+\ii\eps\alpha}{w'-\ii\eps\alpha},\quad
  &&\alpha_2 = \frac{\alpha^2+\ii w'\eps\alpha}{1-\ii w'\eps\alpha}\\
  &\alpha_3 = \frac{w\alpha^2-\alpha}{w-\alpha},\quad
  &&\alpha_4 = \frac{\alpha^2-w\alpha}{1-w\alpha},
 \end{aligned}
\end{equation}
where $\alpha:=\sqrt{-\alpha_1\alpha_2}=\sqrt{\alpha_3\alpha_4}$, $\alpha\bar\alpha=1$ and $\eps=\pm1$. Note that $f_n$ and $\Omega^{(n)}$ ($n=1,\dots,4$) can be expressed in terms of the parameters $\alpha_i$, $K_i$ ($i=1,\dots,4$) via \eqref{genf} and \eqref{kapom} or \eqref{B2} 
and therefore as functions of $w$, $w'$ and $\alpha$ via \eqref{alpha1}.

\subsection{Point-like horizons}
For point-like horizons $K_1=K_2$ or/and $K_3=K_4$ the parametrization \eqref{albet1} does not apply since the mapping of the four coefficients $q$, $r$, $s$, $t$ in \eqref{fp} onto less then four parameters $\alpha_i$ ($\alpha_1=\alpha_2$ or/and $\alpha_3=\alpha_4$) is not invertible. However, the invertibility can be restored by introducing the \emph{derivatives} $\alpha'(\zW)$, $\beta'(\zW)$ in the confluent points $K_1=K_2$ or/and $K_3=K_4$. Let us illustrate the procedure by means of \emph{one} confluent point, say $K_1=K_2$.

Differentiating Eqs. \eqref{albet} at $\zW=K_1$ one obtains
\begin{eqnarray}\label{alprime}\label{d2eq}
 d_2(K_1)\alpha'(K_1) & = & \bar d_2'(K_1)-\alpha_1\bar d_2'(K_1),\\
 n_2(K_1)\beta'(K_1)  & = & \bar n_2'(K_1)-\beta_1 \bar n_2'(K_1),
\end{eqnarray}
where $\alpha'(K_1)=(\partial\alpha/\partial\zW)|_{\zW=K_1}$ etc.
Since $\zW=K_1=K_2$ is a double zero of $\ee^{2U^+}$, see \eqref{Up}, we have
\begin{equation}
 \ee^{2U^+}\big|_{\zW=K_1}=0,\quad
 (\ee^{2U^+})'\big|_{\zW=K_1}=0
\end{equation}
whence
\begin{equation}
 f(K_1)\equiv f_1=-\bar f_1,\quad
 f'(K_1)=-\bar f'(K_1)
\end{equation}
with the consequences
\begin{equation}\label{alalp}
 \alpha_1=-\beta_1,\quad
 \alpha'(K_1)=-\beta'(K_1).
\end{equation}
Together with the three linear algebraic equations
$\bar d_2(K_i)-\alpha_i d_2(K_i)=0$, ($i=1,3,4$) in \eqref{lineq}, Eq. \eqref{alprime} forms an algebraic system consisting of four linear equations for the calculation of $s$, $\bar s$, $t$, $\bar t$ and finally $d_2(\zW)$ in terms of $\alpha_i$, $K_i$ ($i=1,3,4$) and $\alpha'(K_1)$. According to \eqref{alalp}, $n_2(\zW)$ can simply be read off from $d_2(\zW)$ by replacing $\alpha_i$ by $-\alpha_i$ ($i=1,3,4$) and $\alpha'(K_1)$ by $-\alpha'(K_1)$. In this way, we obtain a parameter representation for the axis values $f^+(\zW)$ of the Ernst potential \eqref{fp}. The continuation of $f^+$ to all space can be made easier by the following consideration.

Obviously, the first set of equations in \eqref{lineq} can be reformulated by replacing its second equation by the difference of the second and the first equation,
\begin{equation}\label{diffeq}
 d_2(K_2)\frac{\alpha_2-\alpha_1}{K_2-K_1}=\frac{\bar d_2(K_2)-\bar d_2(K_1)}{K_2-K_1}-\alpha_1\frac{d_2(K_2)-d_2(K_1)}{K_2-K_1},
\end{equation}
and leaving the other three equations ($i=1,3,4$) unchanged. As a consequence, we change the parametrization in the solution of the linear algebraic system for $s$, $r$, $\bar s$, $\bar t$ by replacing $\alpha_2$ by
\begin{equation}\label{subs}
 \alpha_2=\alpha_{21}(K_2-K_1)+\alpha_1,\quad
 \alpha_{21}:=\frac{\alpha_2-\alpha_1}{K_2-K_1}.
\end{equation}
Comparing now equations \eqref{diffeq} and \eqref{d2eq} we find an easy procedure to ``construct'' the gravitational potentials for point-like horizons from the corresponding potentials for the extended case: One has simply to substitute $\alpha_2$ according to \eqref{subs} and to perform the formal limit
\begin{equation}
 K_2\to K_1,\quad \alpha_2\to\alpha_1,\quad \alpha_{21}\to\alpha'(K_1).
\end{equation}
In an analogous way one can introduce $\alpha_{43}$ and, if required, replace it by $\alpha'(K_3)$,
\begin{equation}
 K_4\to K_3,\quad \alpha_4\to\alpha_3,\quad\alpha_{43}\to\alpha'(K_3).
\end{equation}
In this way we reformulate $f(\rW,\zW)$ in \eqref{Yamazaki},
\begin{eqnarray}
 R_{12} & = & \frac{(\alpha_1+\alpha_2)(K_1+K_2-2\zW)}{2(r_1+r_2)}
	      +\frac{\alpha_{21}(r_1+r_2)}{2}\\
  R_{34} & = & \frac{(\alpha_3+\alpha_4)(K_3+K_4-2\zW)}{2(r_3+r_4)}
	      +\frac{\alpha_{43}(r_3+r_4)}{2}\\
 R_{14} & = & \frac{\alpha_1 r_1-\alpha_4 r_4}{K_{14}},\quad
 R_{23} =  \frac{\alpha_2 r_2-\alpha_3 r_3}{K_{23}}.
\end{eqnarray}
For $K_1>K_2>K_3>K_4$ these formulae describe the extended case and are completely equivalent to \eqref{Yamazaki}. For a point-like horizon at $\zW=K_1=K_2$ and an extended horizon $K_2>K_3>K_4$ we have ($K_2\to K_1$, $\alpha_2\to\alpha_1$, $\alpha_{21}\to \alpha'(K_1)$)
\begin{equation}
 \begin{aligned}
 &R_{12} =  \frac{\alpha_1(K_1-\zW)}{r_1}+\alpha'(K_1) r_1,\quad
 &&R_{34} = \frac{\alpha_3 r_3-\alpha_4 r_4}{K_{34}},\\
 &R_{14} =  \frac{\alpha_1 r_1-\alpha_4 r_4}{K_{14}},\quad
 &&R_{23} = \frac{\alpha_1 r_1-\alpha_3 r_3}{K_{13}}.
 \end{aligned}
\end{equation}

Finally, two point-like horizons at $\zW=K_1=K_2$ and $\zW=K_3=K_4$ are described by
\begin{equation}
 \begin{aligned}
  &R_{12} =  \frac{\alpha_1(K_1-\zW)}{r_1}+\alpha'(K_1) r_1,\quad
  &&R_{14} =  \frac{\alpha_1 r_1-\alpha_3 r_3}{K_{13}},\\
  &R_{34} =  \frac{\alpha_1(K_3-\zW)}{r_3}+\alpha'(K_3) r_3,\quad
  &&R_{23} =  \frac{\alpha_1 r_1-\alpha_3 r_3}{K_{13}}.
 \end{aligned}
\end{equation}
Note that $\alpha'(K_1)$ and $\alpha'(K_3)$ can be replaced by the real constants $\gamma_1$ and $\gamma_3$,
\begin{equation}
 \gamma_1=\bar\gamma_1=\ii\frac{\alpha'(K_1)}{\alpha_1},\quad
 \gamma_3=\bar\gamma_3=\ii\frac{\alpha'(K_3)}{\alpha_3}.
\end{equation}

According to \eqref{a} and \eqref{k1}, one obtains the gravitational potentials $a(\rW,\zW)$ and $k(\rW,\zW)$ via line integrals\footnote{For technical details see \cite{Kramer1986}.}. In a first step, one has to perform the integration in the extended case and reparametrize the results in terms of $\alpha_1$, $\alpha_3$, $\alpha_{21}$ and $\alpha_{43}$ afterwards. In a second step one determines the axis values $k(\rW=0,\zW)$, $a(\rW=0,\zW)$ and requires regularity according to \eqref{B1}. In this way one gets from $k=0$ on $\mathcal A^\pm$, $\An$ the relation\footnote{Note that the condition $k=0$ has a second solution  which, however, turns out not to be compatible with the boundary condition $a=0$.}
\begin{equation}\label{kbed}
 \alpha_1\alpha_2+\alpha_3\alpha_4=0
\end{equation}
and from $a=0$ on $\mathcal A^\pm$, $\An$ and \eqref{kbed} the two further relations,
\begin{equation}\label{abed}
 \begin{aligned}
 \alpha_3(1-\alpha_4)^2(K_{41}+K_{32})-(1-\alpha_3\alpha_4)\alpha_{43}K_{31}K_{32}=0,\\
 \alpha_1(1+\alpha_2)^2(K_{41}+K_{32})+(1-\alpha_1\alpha_2)\alpha_{21}K_{31}K_{41}=0.
 \end{aligned}
\end{equation}
Again, these equilibrium conditions contain the ``extended case'' $K_1>K_2>K_3>K_4$, $K_{12}\neq0\neq K_{34}$ as well as the cases with point-like horizons $K_{12}=0$ or/and $K_{34}=0$. Indeed, \eqref{kbed} is identical with \eqref{Bed1} and \eqref{abed} can easily be reformulated to take the form \eqref{Bed2} provided that $K_{12}\neq 0\neq K_{34}$. To describe point-like configurations one has simply to set
\begin{equation}\label{ers}
 \begin{aligned}
  &K_2=K_1,\quad \alpha_2=\alpha_1,\quad\alpha_{21}=\alpha'(K_1)=-\ii\alpha_1\gamma_1\quad
  \textrm{or/and}\\
  &K_4=K_3,\quad \alpha_4=\alpha_3,\quad\alpha_{43}=\alpha'(K_3)=-\ii\alpha_3\gamma_3.
 \end{aligned}
\end{equation}
The subsequent discussions are based on \eqref{kbed}-\eqref{ers}. We will show that, in the degenerate case, these equilibrium conditions have solutions which can be obtained as a limit of equilibrium solutions for extended horizons (as described by \eqref{Bed1}, \eqref{Bed2}). On the other hand, we will find new solution branches for point-like horizons which have no counterpart in the non-degenerate case.

\subsubsection{Configurations with one point-like and one extended horizon\label{PE}}

Without loss of generality, we may assume that the upper horizon be
point-like ($K_1=K_2$) and the lower one be extended ($K_2>K_3>K_4$), see Fig.~\ref{fig:1}b. In
this situation, the equilibrium condition \eqref{kbed} becomes
\begin{equation}\label{al1}
 \alpha_1^2+\alpha_3\alpha_4=0
\end{equation}
with the solution
\begin{equation}
 \alpha_1=\ii\eps\alpha,\quad
 \alpha_3\alpha_4=\alpha^2,
\end{equation}
where $\eps=\pm1$ and $\alpha\in\C$, $\alpha\bar\alpha=1$.
Together with this result, the conditions in \eqref{abed} reduce to
\begin{equation}\label{al4}
 (1-\alpha_4)w=\frac{\alpha^2-\alpha_4}{\alpha},\quad
 (\alpha-\ii\eps)\left[\gamma_1 K_{23}w(\eps\alpha+\ii)+\ii(w+1)(\eps\alpha-\ii)\right]=0.
\end{equation}
The second equation is satisfied if either the first or the second bracket vanishes. Hence,
we obtain two different solutions of the equilibrium
conditions. In the first solution branch ($\alpha\neq\ii\eps$), the parameters $\alpha_1$,
$\gamma_1$, $\alpha_3$ and $\alpha_4$ have to be chosen according to
\begin{equation}\label{L1}
 \alpha_1=\ii\eps\alpha,\quad
 \gamma_1=\frac{\ii(w+1)}{w K_{23}}\frac{\ii-\eps\alpha}{\ii+\eps\alpha},\quad
 \alpha_3=\frac{w\alpha^2-\alpha}{w-\alpha},\quad
 \alpha_4=\frac{\alpha^2-w\alpha}{1-w\alpha}.
\end{equation}
This family of
solutions depends on the two physical parameters $\alpha$ and $w$ (and
on two additional scaling parameters, e.g. $K_1$ and $K_{23}$).
Together with \eqref{ers} it follows that
this solution is nothing but the limit $K_1\to K_2$ ($\Leftrightarrow
w'\to1$) of the solution \eqref{alpha1} for extended horizons.

The second solution branch of the equilibrium conditions is given by
\begin{equation}\label{L2}
 \alpha_1=-1,\quad
 \gamma_1\in\R,\quad
 \alpha_3=\frac{1-\ii\eps w}{1+\ii\eps w},\quad
 \alpha_4=-\frac{1+\ii\eps w}{1-\ii\eps w},
\end{equation}
i.e. the corresponding Ernst potential depends
on the two parameters $\gamma_1$ and $w$
(plus two scaling parameters).
Interestingly, this solution has no counterpart in the case of extended
horizons. It appears only in the present situation with $K_1=K_2$, $K_3\neq K_4$.

\subsubsection{Configurations with two point-like horizons}

In the case of two point-like horizons (cf. Fig.~\ref{fig:1}d), the
equilibrium condition \eqref{kbed}
leads to
\begin{equation}\label{al2}
 \alpha_1^2+\alpha_3^2=0
\end{equation}
which can be solved by setting
\begin{equation}
 \alpha_1=\ii\eps\alpha,\quad
 \alpha_3=-\alpha,
\end{equation}
with $\eps=\pm1$ and $\alpha\in\C$, $\alpha\bar\alpha=1$.
Plugging this into \eqref{abed}, we obtain the two further constraints 
\begin{equation}\label{al3}
 (\alpha+1)\left[(\alpha-1)\gamma_3K_{23}-2\ii(\alpha+1)\right]=0,\quad
 (\alpha-\ii\eps)\left[\gamma_1 K_{23}(\eps\alpha+\ii)+2\ii(\eps\alpha-\ii)\right]=0.
\end{equation}
Note that the second equation in \eqref{al3} is just the limit $w\to1$ of the second equation in \eqref{al4}.

By solving \eqref{al3}, we obtain three different solution branches for configurations with two point-like horizons.
The first one
\begin{equation}\label{L3}
 \alpha_1=\ii\eps\alpha,\quad \alpha_3=-\alpha,\quad
 \gamma_1=\frac{2\ii}{K_{23}}\cdot\frac{1+\ii\eps\alpha}{1-\ii\eps\alpha},
 \quad
 \gamma_3=\frac{2\ii}{K_{23}}\cdot\frac{\alpha+1}{\alpha-1},
\end{equation}
depends on one free parameter $\alpha$. This solution can be obtained in the limit
$K_1\to K_2$, $K_3\to K_4$ ($\Leftrightarrow w\to1,w'\to 1$) from
\eqref{alpha1}.

The second and third solution branches are
\begin{equation}\label{L4}
 \alpha_1=-\ii\eps,\quad \alpha_3=1, \quad
 \gamma_1=\frac{2\eps}{K_{23}},
 \quad
 \gamma_3\in\R,
\end{equation}
and
\begin{equation}\label{L5}
 \alpha_1=-1,\quad \alpha_3=-\ii\eps,\quad \gamma_1\in\R,\quad
 \gamma_3=\frac{2\eps}{ K_{23}},
\end{equation}
where now $\gamma_3$ or $\gamma_1$ are free parameters.
It turns out that \eqref{L4} and \eqref{L5} describe the same
physical situation --- the positions of the two degenerate objects are
merely interchanged (i.e. both solutions differ only by a coordinate
transformation).  Therefore, it is sufficient to study the solutions
branches \eqref{L3} and \eqref{L4}.

\section{Black hole inequalities and singularities}
\subsection{Sub-extremal black holes}

In paper I we have analysed the possible equilibrium between two \emph{sub-extremal} black holes. Following Both and Fairhurst \cite{Booth}, we have defined sub-extremality through existence of trapped surfaces (surfaces with a negative expansion of outgoing null rays) in every sufficiently small interior neighbourhood of the event horizon. It can be shown \cite{Hennig2008a} that any such axisymmetric and stationary sub-extremal black hole satisfies the inequality
\begin{equation}\label{ineq}
 8\pi|J|<A
\end{equation}
between angular momentum $J$ and horizon area $A$. This inequality was the key ingredient for the non-existence proof in paper I: Using the explicit expressions for the angular momenta $J_1$, $J_2$ and the horizon areas $A_1$, $A_2$ of the two gravitational sources described by the two-horizon solution, it followed that at least one of these objects violates $8\pi|J_i|<A_i$, $i=1,2$. This proves that a regular equilibrium configuration containing two sub-extremal black holes does not exist.
\subsection{Degenerate black holes\label{sec:deg}}

A degenerate black hole is defined by vanishing surface gravity,
\begin{equation}\label{kappa}
 \kappa=0.
\end{equation}
As shown by Ansorg and Pfister \cite{Ansorg2008a}, such black holes
satisfy the universal relation\footnote{The theorem in
\cite{Ansorg2008a} makes originally
two assumptions (namely equatorial symmetry and
existence of a continuous sequence of spacetimes, leading from the
Kerr-Newman solution in electrovacuum to the discussed black hole
solutions) that turn out to be not necessary and therefore
can be dropped, see Appendix A in \cite{Hennig2008b}.} 
\begin{equation}\label{eq}
8\pi|J|=A
\end{equation}
instead of inequality \eqref{ineq}.

From a discussion of
the expressions for $\kappa_1$, $\kappa_2$, $J_1$, $J_2$, $A_1$ and
$A_2$ for our solution of the two-black-hole boundary value problem (the
formulae for these ``thermodynamic quantities''
can be found in paper I) it
follows that the upper gravitational object is degenerate (i.e. satisfies
\eqref{kappa} and \eqref{eq}) if and only if $w'=1$ holds. Similarly,
the lower object is degenerate if and only if $w=1$ holds. According
to the equivalences
\begin{equation}
 w'=1\quad\Leftrightarrow\quad K_1=K_2\quad\textrm{and}\quad
 w =1\quad\Leftrightarrow\quad K_3=K_4,
\end{equation}
which follow from \eqref{Bed2}, we see that
 degenerate black holes in a possible two-black-hole equilibrium
configuration are described precisely by the earlier discussed ``point-like''
horizons.

Now we have all the ingredients to perform the desired non-existence
proof for two-black hole configurations with at least one degenerate
black hole. 

\subsubsection{One degenerate and one sub-extremal black hole}

As before, we can restrict ourselves to the situation as sketched in
Fig.~\ref{fig:1}b, i.e. we assume that the upper black hole is
degenerate and the lower one is sub-extremal.  The opposite
configuration in Fig.~\ref{fig:1}c differs from this situation
only by a coordinate change.

As shown in Sec.~\ref{PE}, there are two different
families of solutions that are
candidates for this equilibrium situation, which we study separately in
the following.

We start by considering the solution \eqref{L1}. Since the lower black
hole is assumed to be sub-extremal, it has to satisfy 
$8\pi|J_2|<A_2$.  The latter inequality is equivalent to $1-p_2^2>0$,
where $p_2:=8\pi J_2/A_2$. Calculating area and angular momentum with
the formulae in paper I, we find
\begin{equation}
p_2=\eps\frac{w(w-\Phi)}{1-w\Phi},
\end{equation}
where
\begin{equation}\label{Phi}
 \Phi:=\cos\phi+\eps\sin\phi\equiv\sqrt{2}
       \sin\left(x+\frac{\pi}{4}\right),\quad x:=\eps\phi,\quad
 \alpha=:\ee^{\ii\phi}.
\end{equation}
This leads to
\begin{equation}
 1-p_2^2 = -(w^2-2\Phi w+1)\frac{w^2-1}{(w\Phi-1)^2}>0.
\end{equation}
Using $w\in(1,\infty)$ we conclude
\begin{equation}\label{ineq2}
 w^2-2\Phi w+1<0.
\end{equation}
Now we define
\begin{equation}
 \omega:=\frac{1}{2}\left(w+\frac{1}{w}\right)
\end{equation}
and use \eqref{ineq2} and \eqref{Phi} to obtain
\begin{equation}
 1<\omega<\Phi\le \sqrt{2}.
\end{equation}
The inequality $\Phi>1$ leads to the restriction
\begin{equation}
 x\in\left(0,\frac{\pi}{2}\right)\big|_{\textrm{mod }2\pi}
\end{equation}
for the parameter $x$. In particular, this implies
\begin{equation}
 \sin x>0, \quad\cos x>0.
\end{equation}

Now we use these results in order to estimate 
the ADM mass $M$ of the spacetime. The explicit calculation shows that
$M$ can be written in terms of $x$ and $\omega$ as
\begin{equation}
  M=-\frac{K_{23}}{2}(1+w)\frac{1+\sin x - \omega\cos x}
     {1+\sin x-\omega\cos x+\sin x\cos x},
\end{equation}
where $K_{23}$ is positive since we consider two \emph{separated} horizons.
With $\omega<\Phi$, we can estimate
\begin{eqnarray}
 1+\sin x-\omega\cos(x) & > &\nonumber
    1+\sin x-(\sin x+\cos x)\cos x\\ \nonumber
 & = & \sin x\,(1+\sin x-\cos x)\\ \nonumber
 & = & \sin x\,\left[1+\sqrt{2}
        \sin\left(x-\frac{\pi}{4}\right)\right]\\
 & > & 0 \quad\textrm{for}\quad
         x\in\left(0,\frac{\pi}{2}\right).
\end{eqnarray}
Thus we arrive at $M<0$ in contradiction to the positive
mass theorem. 

Similarly, we study the second solution branch \eqref{L2}. Here,
we obtain
\begin{equation}
 p_2\equiv\frac{8\pi J_2}{A_2}
 =\frac{\gamma_1 K_{23}(1+w^2)w}{2(\eps\gamma_1 K_{23}w^2+1-w)}
\end{equation}
and
\begin{equation}
 1-p_2^2=-\frac{1}{4}(w-1)\frac{[\eps\gamma_1 K_{23}(w-1)w+2]
          [\eps\gamma_1 K_{23}(1+w)^2w-2(w-1)]}
          {(\eps\gamma_1 K_{23}w^2+1-w)^2}.
\end{equation}
The inequality $8\pi|J_2|<A_2$ for the sub-extremal black hole
(in the form $1-p_2^2>0$) implies
\begin{equation}\label{I1}
 -\frac{2}{w(w-1)}<\eps\gamma_1 K_{23}<\frac{2(w-1)}{w(1+w)^2}
\end{equation}
for the parameters $w$ and $\gamma_1$. Now we study the ADM-mass for
this solution branch. We obtain
\begin{equation}
 M=-\frac{K_{23}}{2}\cdot\frac{\eps\gamma_1 K_{23}(1+w^2)-2(1+w)}
    {\eps\gamma_1 K_{23}(1+w)-2}.
\end{equation}
From this expression it follows that $M$ is positive only for
\begin{equation}\label{I2}
 \frac{2}{1+w}<\eps\gamma_1 K_{23}<\frac{2(1+w)}{1+w^2}.
\end{equation}
Due to
\begin{equation}
 \frac{2(w-1)}{w(1+w)^2}<\frac{2}{(1+w)^2}<\frac{2}{1+w}
\end{equation}
for the allowed w-values $w\in(1,\infty)$, the interval in 
\eqref{I2} is always disjunct from the interval \eqref{I1}. Hence, the
ADM mass is negative for this solution branch, too.

Therefore, we conclude that configurations with one degenerate and one
sub-extremal black hole cannot be in equilibrium. 

\subsubsection{Configurations with two degenerate black holes}

In the situation with two degenerate objects, the equilibrium conditions
have the solution families \eqref{L3}, \eqref{L4} and \eqref{L5} ---
but, as mentioned earlier, \eqref{L5} is physically
equivalent to \eqref{L4} and needs therefore not to be treated
separately here.

For the solution branch \eqref{L3}, the ADM mass is given by
\begin{equation}
 M=-\frac{2K_{23}}{3+\sqrt{2}\cos
    \left(\eps\phi+\frac{\pi}{4}\right)},\quad
 \alpha=:\ee^{\ii\phi},
\end{equation}
which is obviously negative, such that this branch can be
excluded immediately. 

In the second case \eqref{L4}, we obtain
\begin{equation}
 M=-\frac{K_{23}}{2}\cdot\frac{\tilde\gamma_3-2}{\tilde\gamma_3-1},\quad
 \tilde\gamma_3:=\eps \gamma_3 K_{23}.
\end{equation}
Obviously, $M$ is negative for $\tilde\gamma_3<1$ and for
$\tilde\gamma_3>2$, such that these parameter regions can be
excluded due to the positive mass theorem. However, $M\ge 0$ holds in
the range $\tilde\gamma_3\in[1,2]$.
As a next step, it is interesting to test
whether the Penrose inequality, i.e. the
stronger condition
\begin{equation}
M>\sqrt{\frac{A_1+A_2}{16\pi}}
\end{equation}
is satisfied in this
parameter range (even though it is not yet proved that the Penrose
inequality holds for general axisymmetric and stationary black hole
spacetimes). With
\begin{equation}
 A_1=\frac{2\pi\tilde\gamma_3^2 K_{23}^2}{(\tilde\gamma_3-1)^2},\quad
 A_2=\frac{4\pi K_{23}^2}{(\tilde\gamma_3-1)^2}
\end{equation}
we arrive at
\begin{equation}
 16\pi M^2-A_1-A_2=\frac{2\pi K_{23}^2}{(\tilde\gamma_3-1)^2}
                    (\tilde\gamma_3^2-8\tilde\gamma_3+6)
 <0\quad\textrm{for}\quad\tilde\gamma_3\in[1,2],
\end{equation}
i.e. we find a violation of the Penrose inequality. Since this
inequality is related to cosmic censorship, such a violation might indicate
that the  investigated spacetimes do not possess a regular exterior
vacuum region outside the two gravitational sources. 
And this is indeed what we observe: It can be shown (see Appendix~\ref{App})
that there are always \emph{singular rings} at
which the Ernst potential $f$ diverges. Since $f$
can be defined invariantly in terms of the two Killing vectors, this behaviour is not just a coordinate effect but results from a physical singularity. Therefore, this solution branch can be excluded, too.

\section{Discussion}

We have studied the question whether two-black-hole configurations containing degenerate black holes can be in stationary equilibrium. This question can be discussed in terms of a boundary value problem for the Einstein equations. Using the complex Ernst formulation and the inverse scattering method we have solved this boundary value problem and shown that particular degenerate limits of the double-Kerr-NUT solution are the only candidates for the desired equilibrium configurations. However, as a careful discussion of these solutions has revealed, they all contain singularities outside the black holes and therefore do not represent physically acceptable black hole configurations. Together with the corresponding result for non-degenerate two-black-hole configurations in \cite{Neugebauer2009} (paper I) we have shown that axisymmetric and stationary configurations containing
\begin{itemize}
 \item two sub-extremal black holes (defined by existence of trapped
       surfaces just inside the event horizon), or
 \item two degenerate black holes (defined by vanishing surface
       gravity), or
 \item one degenerate and one sub-extremal black hole
\end{itemize}
cannot be in equilibrium.


\begin{acknowledgements}
We would like to thank Reinhard Meinel for interesting discussions and Ben Whale for commenting on the manuscript.
\end{acknowledgements}
\begin{appendix}
\section{Singular Ernst potential\label{App}}
In this appendix we show that the second solution branch \eqref{L4} for
configurations with two degenerate horizons has always singularities
outside the two gravitational objects.
For that purpose, we introduce dimensionless coordinates $\tilde\rW$ and
$\tilde\zeta$ by
\begin{equation}
 \rW = K_{23}\tilde\rW,\quad \zeta=K_1+K_{23}\tilde\zeta.
\end{equation}
The Ernst potential $f$ can then be written as
\begin{equation}\label{f1}
 f=\frac{A_+}{A_-}
\end{equation}
with
\begin{eqnarray}
 A_\pm & := & -\frac{1}{2}\left[W_2^2(W_1^2-W_2^2+1)\tilde\gamma_3
              +W_1(2W_1-2W_2\pm1)(W_1^2-(W_2\pm1)^2)\right]\nonumber\\
       &&     +\frac{\ii\eps}{4}\left[4W_1W_2^2(2W_1\pm1)\tilde\gamma_3
              -(W_1^4+W_2^4\pm2W_2^3\mp2W_2-1
              +6W_1^2W_2^2\pm6W_1^2W_2)\right],   
\end{eqnarray}
where
\begin{equation}\label{WW}
 W_1:=\sqrt{\tilde\rW^2+\tilde\zeta^2},\quad
 W_2:=\sqrt{\tilde\rW^2+(\tilde\zeta+1)^2},\quad
 \tilde\gamma_3:=\eps K_{23}\gamma_3.
\end{equation}

The Ernst potential becomes singular if there are (real) values for
$\tilde\rW$ and $\tilde\zeta$ for which $A_-=0$ holds. (It can be shown
that $A_+$ and $A_-$ cannot vanish simultaneously, i.e. the
numerator in \eqref{f1} cannot compensate zeros of the denominator.)
In the following we show that such $\tilde\rW$-$\tilde\zeta$-values
indeed exist for all $\tilde\gamma_3\in[1,2]$ (i.e. in the entire range of
positive ADM mass\footnote{Actually, there is numerical evidence that the
Ernst potential of the double-Kerr-NUT solution suffers \emph{always} from the
presence of singular rings whenever the equilibrium conditions are
satisfied (and not only for the configurations with two degenerate black
holes that are discussed here). However, a rigorous proof of this
statement seems to be difficult.}). For that purpose, we solve the two
equations $\Re A_-=0$, $\Im A_-=0$ explicitly:
\begin{eqnarray}\label{S1a}
 W_2^2(W_2^2-W_1^2-1)\tilde\gamma_3 & = & 
              W_1(2W_1-2W_2-1)(W_1^2-(W_2-1)^2),\\
 \label{S1b}
 4W_1W_2^2(2W_1-1)\tilde\gamma_3 & = & 
              W_1^4+W_2^4-2W_2^3+2W_2-1
              +6W_1^2W_2^2-6W_1^2W_2.             
\end{eqnarray}

It turns out that these equations can be simplified by replacing
$W_1$, $W_2$ and $\tilde\gamma_3$ via
\begin{equation}\label{W}
 W_1=(v_1v_2-1)(1+v_1),\quad
 W_2=(v_1v_2-1)(1-v_1)+1,\quad
 \tilde\gamma_3=\frac{2(v_1v_2-1)^2}{v_1(v_2-v_1v_2+1)^2}\gamma
\end{equation}
in terms of new variables $v_1$, $v_2$ and $\gamma$.
Then, the system \eqref{S1a}, \eqref{S1b} becomes
\begin{eqnarray}
 \label{S2a}
 (v_1-2v_1^2v_2+1)\gamma & = & (1+v_1)(4v_1^2v_2-4v_1-3),\\
 \label{S2b}
 (1+v_1)(2v_1^2v_2+2v_1v_2-2v_1-3)\gamma & = & v_1^4v_2-v_1^3-v_1^2+v_2.
\end{eqnarray}
The advantage of this formulation is that now both equations are linear
in $v_2$. 

We can combine \eqref{S2a} and \eqref{S2b}, on the one hand, such that
$v_2$ is eliminated and, on the other hand, such that $\gamma$ is
eliminated. In this way, we obtain
\begin{eqnarray}
 \label{S3a}
 &&(1+\gamma)v_1^4-2\gamma(\gamma-1)v_1^3-2\gamma(\gamma-4)v_1^2
 +2(\gamma^2+3\gamma-2)v_1-\gamma-3=0,\\
 &&2v_1^2(5v_1^4+12v_1^3+12v_1^2+4v_1+1)v_2^2\nonumber\\
 \label{S3b}
 &&\qquad-(1+v_1)(19v_1^4+42v_1^3+32v_1^2+6v_1+1)v_2+9(1+v_1)^4=0,  
\end{eqnarray}
i.e. we arrive at a quartic equation for $v_1=v_1(\gamma)$ and a
quadratic equation from which $v_2$ can be calculated from $v_1$.
Since, fortunately, fourth order equations still belong to the class of
completely solvable algebraic equations, we are able to find an
explicit solution.

A particular solution of the system \eqref{S3a}, \eqref{S3b} that
corresponds to real values for $\tilde\rW$, $\tilde\zeta$ is given by
\begin{eqnarray}
 v_1 & = & \frac{1}{2(\gamma+1)}\left[\gamma(\gamma-1)-T_3
           +\sqrt{\frac{3P_3}{2}-\frac{2T_2}{3}-\frac{2P_4}{T_3}}\,\right],\\
 v_2 & = & \frac{(1+v_1)\left[19v_1^4+42v_1^3+32v_1^2+6v_1+1-(1+v_1^2)
           \sqrt{v_1^4+12v_1^3+26v_1^2+12v_1+1}\right]}
           {4v_1^2(5v_1^4+12v_1^3+12v_1^2+4v_1+1)},
\end{eqnarray}
where we have used the following auxiliary quantities:
\begin{eqnarray}
 P_1 & := & 4\gamma^4-2\gamma^3-2\gamma^2-6\gamma-9,\\
 P_2 & := & 11\gamma^6-42\gamma^5-84\gamma^4-146\gamma^3-171\gamma^2
            -108\gamma-54,\\
 P_3 & := & \gamma(3\gamma^3-2\gamma^2-9\gamma-16),\\
 P_4 & := & \gamma^6-\gamma^5-7\gamma^4-13\gamma^3-2\gamma^2+2\gamma+4,\\
 T_1 & := & \left(\frac{P_2}{2}
            +\sqrt{\frac{P_2^2}{4}-P_1^3}\right)^\frac{1}{3},\\
 T_2 & := & \frac{5P_3}{4}-(1+\gamma)\left(\frac{P_1}{T_1}+T_1\right),\\
 T_3 & := & \sqrt{\frac{2T_2}{3}-\frac{P_3}{2}}.
\end{eqnarray}
From the above solution $v_1(\gamma)$, $v_2(\gamma)$ we can calculate
$W_1$, $W_2$ and $\tilde\gamma_3$ using \eqref{W} and, afterwards,
$\tilde\rW$ and $\tilde\zeta$ via
\begin{equation}
 \tilde\rW=\frac{1}{2}\sqrt{4W_1^2-(W_2^2-W_1^2-1)^2},\quad
 \tilde\zeta=\frac{1}{2}(W_2^2-W_1^2-1),
\end{equation}
see \eqref{WW}.

\begin{figure}\centering
 \includegraphics[scale=0.8]{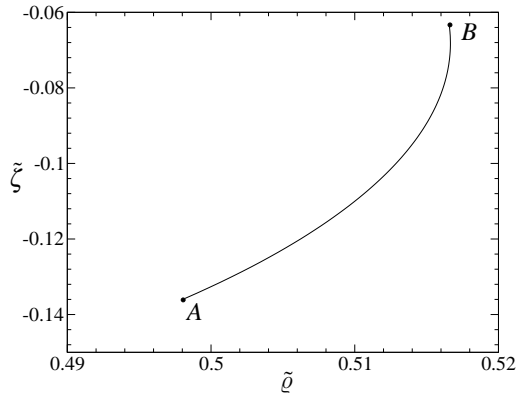}
 \caption{Positions of the singularities of the Ernst potential for
 all values of the parameter $\tilde\gamma_3\in[1,2]$ (the parameter
 region with non-negative ADM mass). The point $A$
 corresponds to $\tilde\gamma_3=1$ ($\gamma=7.46516\dots$) and at $B$ we
 have $\tilde\gamma_3=2$ ($\gamma=17.19824\dots$).}
 \label{fig:2}
\end{figure}

In this way, we obtain a parametric solution
$(\tilde\rW(\gamma),\tilde\zeta(\gamma),\tilde\gamma_3(\gamma))$ of
the equation system  \eqref{S1a}, \eqref{S1b}.
Here, we consider all values $\gamma\in[7.46516\dots, 17.19824\dots]$
since this interval corresponds to
the parameter region $\tilde\gamma_3\in[1,2]$ with non-negative ADM
mass. A plot of the solution can be found in Fig.~\ref{fig:2}.

From our explicit calculation we conclude that the
Ernst potential has singularities (singular rings) in the entire
parameter range with positive ADM mass.

\end{appendix}

\end{document}